# Review on the Security Threats of Internet of Things

Prajoy Podder
Institute of ICT
Bangladesh University of
Engineering and Technology
Dhaka-1205, Bangladesh.
prajoypodder@gmail.com

M. Rubaiyat Hossain Mondal
Institute of ICT
Bangladesh University of
Engineering and Technology
Dhaka-1205, Bangladesh.
rubaiyat97@yahoo.com

Subrato Bharati
Institute of ICT
Bangladesh University of
Engineering and Technology
Dhaka-1205, Bangladesh.
subratobharati1@gmail.com

Pinto Kumar Paul
Department of CSE
Ranada Prasad Shaha University
Narayanganj, Bangladesh.
pinto.kumar07@gmail.com

## ABSTRACT

Internet of Things (IoT) is being considered as the growth engine for industrial revolution 4.0. The combination of IoT, cloud computing and healthcare can contribute in ensuring well-being of people. One important challenge of IoT network is maintaining privacy and to overcome security threats. This paper provides a systematic review of the security aspects of IoT. Firstly, the application of IoT in industrial and medical service scenarios are described, and the security threats are discussed for the different layers of IoT healthcare architecture. Secondly, different types of existing malware including spyware, viruses, worms, keyloggers, and trojan horses are described in the context of IoT. Thirdly, some of the recent malware attacks such as Mirai, echobot and reaper are discussed. Next, a comparative discussion is presented on the effectiveness of different machine learning algorithms in mitigating the security threats. It is found that the k-nearest neighbor (kNN) machine learning algorithm exhibits excellent accuracy in detecting malware. This paper also reviews different tools for ransomware detection, classification and analysis. Finally, a discussion is presented on the existing security issues, open challenges and possible future scopes in ensuring IoT security.

## 1. INTRODUCTION

Many intelligent systems like gadgets and applications are developing day by day based on advanced technology like the Internet of things (IoT). The usage of IoT is increasing day by day because of its importance. IoT has been recently integrated into many gadgets and applications, to make the system linear to rational. This adoption is increasing day by day because of its importance. IoT technology is also effectively influencing our medical science. The healthcare monitoring system is being developed to ensure emergency services to the patients effectively [1]. Some health application is already developed based on IoT such as emergency notification, continuous glucose monitoring (CGM), and computer-assisted rehabilitation. Those software applications are built to solve different aspects of medical issues. Smartphones are the most crucial part of our daily life and the intelligent application uses the sensor of smartphones. They continuously perceive data from the devices using its sensors and try to monitor the subject's health condition [2]. The whole system needs different types of data from the wards and diagnostics equipment. This is to analyze using data mining and to conclude an efficient result for monitoring and tracking purposes [3, 4]. After that, the intelligent system gains the ability to control health care automatically [5-6].

However, there are some challenges in the integration of IoT technology. Data storage problem, data management problem, exchange of information between devices, security and privacy – these are the main problems that need to be solved first. Cloud computing can be addressed as one of the most effective solutions for all of these problems. A conventional healthcare system is presented in Figure 1 that integrates both IoT and cloud computing in order to provide the facility to access shared medical data and common infrastructure transparently and efficiently.

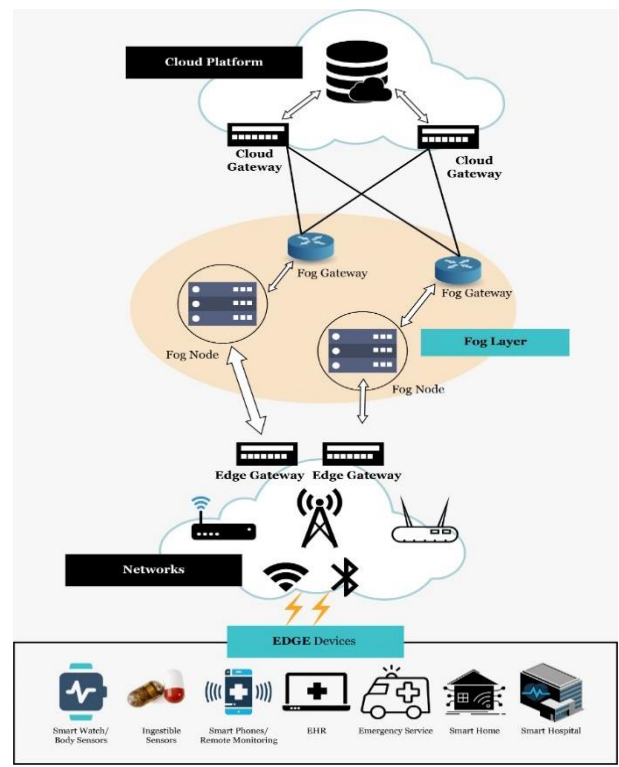

**Fig 1: An overview of cloud integrated IoT system**

Cloud computing offers computing services i.e. software, databases, servers, data analytics, networking over the Internet to deliver faster expansion, economies of scale and flexible resources. In edge devices, fog computing exhibits data analytics, so that it performs real-time processing, reduces costs, and improves the privacy of data. The rise of cloud computing, artificial intelligence and portable devices ensure a solid foundation for the evolution of IoT based healthcare sector.

Medical devices or instruments are also engaged with several wireless communication technologies (i.e., Wi-Fi, bluetooth, etc.) that permit the machine-to-machine communication. It is an environment for Internet of medical things (IoMT) communication. In IoMT, the devices for smart healthcare send data to cloud servers. Several cloud platforms, i.e., Amazon Web Services, may be conducted to store the patients' health related data and to explore the data for health prescriptions and accurate decision making [8].

For rapid deploying and developing the IoT systems, the issues of security in the IoT devices are facing day by day. This increases the probability to launch different types of attacks in the IoMT environment via the Internet. It occurs very crucial issue in the IoMT that controls the smart medical devices with its communication. For example, if an attacker practically obtains the remote control over an IoT based smart medical device, s/he can manipulate the patients' data.

The key motivation behind this work is illustrated as follows. In recent days, IoT devices i.e. smart city, smart home and smart healthcare devices have become the crucial part of our daily life. Since we know it, the users of IoT devices are able to access the data remotely using the Internet [9, 10]. Different entities, such as IoT devices, servers and users, communicate through the Internet. Wi-Fi [11,12], WiMax [13, 14], LTE [15, 16] are the popular forms of using Internet effectively. Light fidelity (LiFi) [11-16] is also an emerging technology for the Internet. OFDM, MIMO systems are playing a vital role in order to establish upgraded wireless communication systems [17-18]. However, there are many forms of security issues in IoT/IoMT communication environment.

The massive scale of IoT based networks carry some new challenges. Existing research papers covered different aspects of IoT such as architecture, communication system, IoT related various applications [19-28], security and privacy [28-29]. However, the mainspring of the commercialization of IoT /IoMT and industrial IoT (IIoT) technology is the security and privacy assurance as well as user satisfaction.

Different sorts of IoT malware are continuously emerging. These can easily affect the communication of IoMT. Malware is also used to control the smart medical devices. Different kinds of attacks, i.e., denial of service, replay, password guessing, impersonation and man-in-the-middle (MITM) attacks can get chance to enter in this environment. Usually, the hackers may apply malware to target the Internet based health care devices for entering illegal access or controlling these devices remotely.

To deploy malware in the environment of IoMT, the hackers adopt network of attacker processes such as botnet. Some examples of botnet are Echobot, Mirai, Emotet, Reaper, Necurs and Gamut. These kinds of botnet attacks are also probable in the environment of IoMT and can be permitted to control or hijack an IoMT based smart healthcare device remotely. This can occur several life threatening conditions for the patients. Consequently, people developing in the IoT security domain emerge with novel ideas to protect the environment of IoMT communication against these attacks. For that reasons, in our work we focus on various types of active IoMT malware and malware programs. The major outcomes of this paper are as follows.

(i) The relation between IoT and cloud computing environment is discussed, and different security requirements of IoT communication environment are illustrated.

(iii) Recent malware attacks such as Mirai, Reaper, Echobot, Emotet, Gamut and Necurs are studied in the context of IoT and IoMT environment.

(iv) The performance of various machine learning techniques for classification and Android malware detection are summarized.

(iv) The effect of ransomware in IoT/IoMT environment is discussed and the existing software tools for ransomware detection are summarized. The remaining part of the manuscript is prepared as follows: Section 2 considers the application of IoT communication environment, while Section 3 discusses security threats. Section 4 introduces various kinds of malware, while Section 5 discusses some of the recent malware. The effectiveness of different machine learning algorithms in recognizing security threats are discussed in Section 6. Ransomware is studied in Section 7, and future research directions are reported in Section 8. The concluding remarks are presented in Section 9.

## 2. APPLICATIONS OF IOT COMMNICATION ENVIRONMENT

In IIoT, sensors and machines in factories, industries are interconnected with each other. Then they provide real-time data over the Internet to the engineers or the manufacturers to increase the industrial processes.

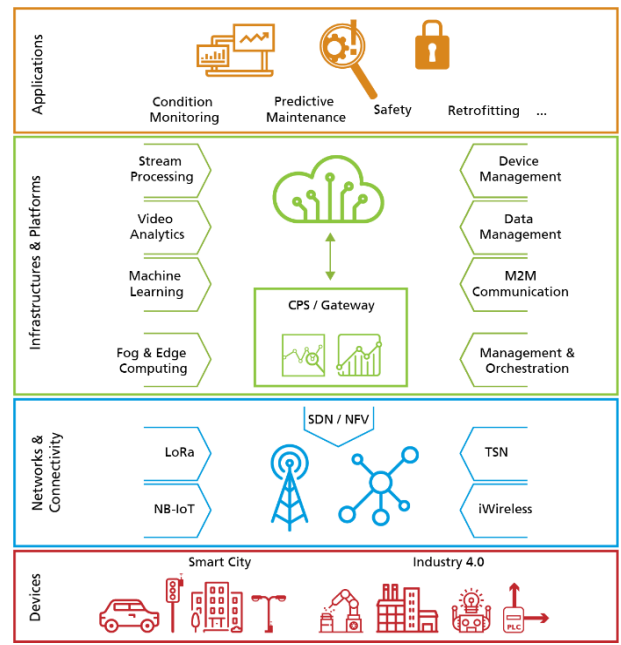

**Fig 2: A general overview of IIoT [61]**

In the medical or healthcare based systems, reactive healthcare based schemes can be changed into proactive wellness-based schemes using IoMT. These types of systems are particular IoMT based smart healthcare or medical devices monitor as well as send medical data to a cloud server. If a patient's relative or a doctor is attracted in the real-time access of this devices, it can be also accomplished by using the environment of IoT. In this way, IoMT facilitates the analysis, processing and access of the suitable medical data.

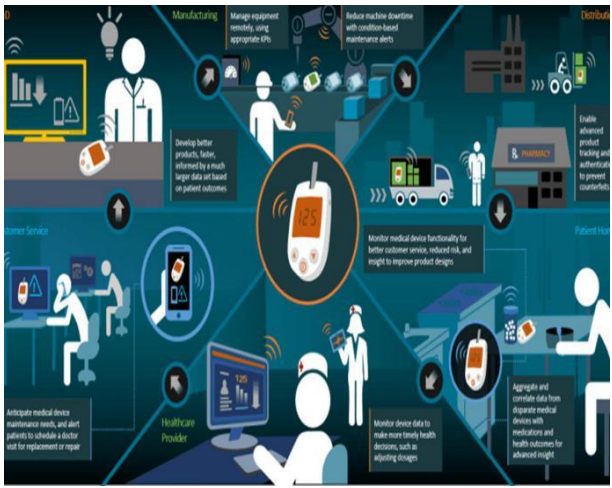

**Fig 3: An overview of IoMT [60]**

## 3. SECURITY AND PRIVACY THREATS IN IOT

Security and privacy threats are summarized in Figure 4 and Figure 5.

The IoT healthcare applications architecture normally consists of the following layers [22]:

a) Application
b) Communication
c) Device
d) Network
e) Transport.

The percentage rate of affecting the layers are as follows:

a) Application layer (9%)
b) Communication layer (18%)
c) Device layer (42%)
d) Network layer (27%)
e) Transport layer (4%).

From the above discussion, it can be said that device layer suffers the highest impact, while the transport layer is affected the lowest.

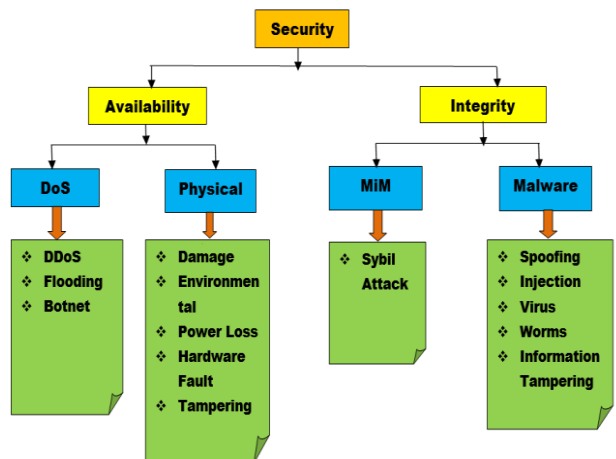

**Fig 4: Security threats of IoT/IoMT**

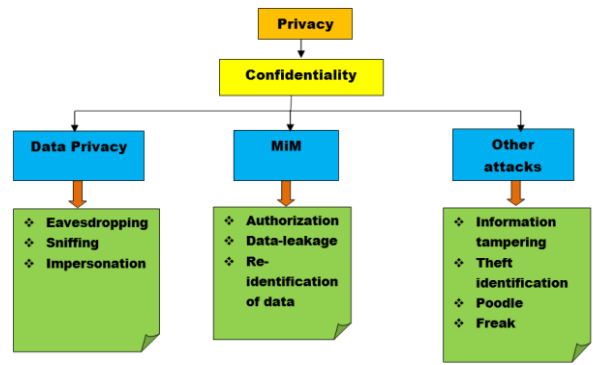

**Fig 5: Privacy threats of IoT/IoMT**

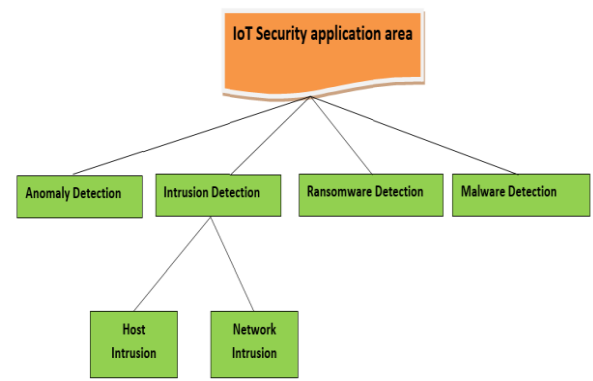

**Fig 6: IoT security application area**

## 4. VARIOUS TYPES OF MALWARE

Malicious software is shortly known as malware in which is a code or program that is generally offered over a network. It conducts or infects various malicious operations that a hacker or an attacker would like to do. Malware can be separated into various categories according to their functionality features.

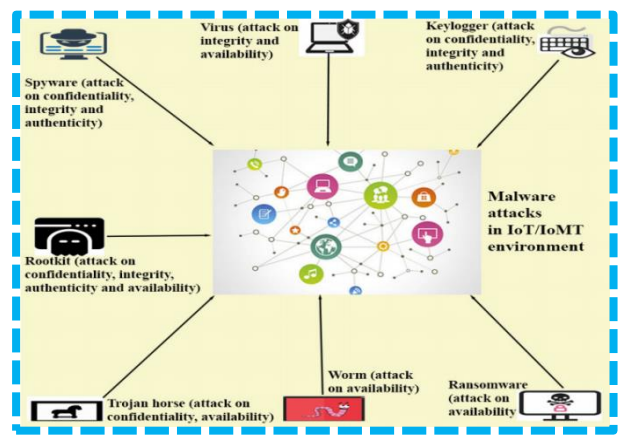

**Fig 7: Different kinds of malware**

Different kinds of malware (shown in Figure 7) are illustrated below [24-29]:

### 4.1 SPYWARE

Spyware is a kind of malware, which involves by spying the active user without their permission. The malicious types of activities like monitoring, collecting keystrokes, harvesting data such as credit card number, financial data, account credentials, are feasible in the network. It may also infect the software security settings in a device. It can take advantage from the vulnerabilities of the free software and then attach itself with several programs.

## 4.2 KEYLOGGER

This type of malicious is a piece of code, which is conducted by a hacker or attacker to track the keystrokes of the operators or users. All information through the keyboard (i.e., their ID, login information, and passwords) have been documented. This malicious attack is stronger than dictionary attack or brute force. The keylogger first attempts to move into a user's Internet based device. It is so terrible that the device cannot be protected with a strong password. Therefore, suggestion for the users is to use multi-factor authentication (i.e., amalgamation of user name, smart card, biometrics data and password).

## 4.3 TROJAN HORSE

This malware pretends itself as a general program to track operators or users into installing after downloading it. In this infected system, it provides the hacker to get opportunity to an authorized remote access. In this system, the hacker can easily steal the data (i.e., credit card information, account number, financial data, password etc.).

## 4.4 VIRUS

This malicious program has a capability of copying itself and deploying to systems. It deploys to each system by including itself to various programs as well as executing the code when a user commences on this infected program. It can steal information, build botnets and damage the host system with the help of this malicious program.

## 4.5 WORMS

It deploys over a network by searching out the weak operating system. It works on the system for damaging their host networks through web servers overload and bandwidth consumption.

## 5. RECENT MALWARE ATTACKS

There are some recent events of malware attacks in IoT/IoMT environment. Some of these are discussed below.

## 5.1 MIRAI

Attacks by Mirai botnet are still going on. Mirai enables monitoring devices running Linux operating systems. These devices can also be conducted as a portion of botnet to carry out different malicious attacks. This malicious program mainly targets smart IoT/IoMT devices, i.e., internet based consumer devices (e.g. home appliances or IP cameras).

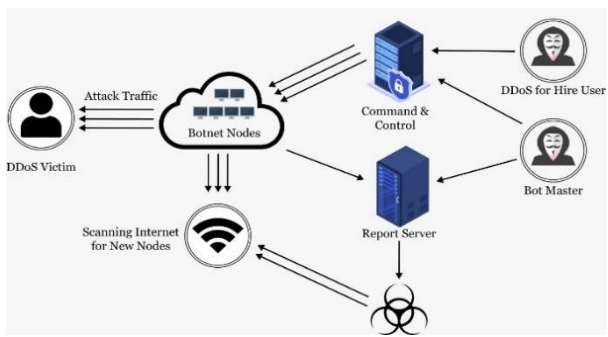

**Fig 8: At a glance of MIRAI**

Mirai was very active botnet along with the statement of Fortinet in 2018. Moreover, this types of botnets have recently extended some features and these are able to infect IoT/IoMT devices. As of Fortinet, Mirai botnets targeted the IoT or IoMT devices for both unknown and known vulnerabilities. In the botnet, crypto mining exhibits up a crucial change. A hacker can conduct the hardware along with electricity of target's scheme to receive the cryptocurrencies via this types of malware. These typical malicious observances are investigating how to conduct IoMT/IoT botnets to create money [52]-[56].

## 5.2 REAPER

Reaper is a malware that is known as IoTroop. Some researchers of information security created this new botnet with enhanced functionality features in 2017. It can compromise with IoT based smart device rapidly as in contrast to the Mirai botnet. Reaper has various effects as it can overthrow the whole infrastructure rapidly. Mirai corrupts the IoT based smart devices which conduct default passwords and user names. Nevertheless, reaper is more terrible which aims 9 different vulnerabilities in various makers' devices i.e. Linksys, Netgear, and D-Link. Employing this botnet, the hacker can change or vary the malware code to make it more destroying. According to the information served by "Recorded Future". It conducted to attack on several EU banks (i.e., ABN Amro) [56]-[58].

## 5.3 ECHOBOT

Echobot is a kind of malware which is the alteration of Mirai. It was revealed in the year of 2019. This types of malware conducts 26 malicious scripts for expanding its activity. Echobot can put out the advantages of unpatched smart IoT based devices [59]. Ecohobot can create huge number of problems for several applications of the enterprise using these vulnerabilities i.e. weblogic of oracle.

## 5.4 EMOTET, GAMUT and NECURS

Emotet, Gamu and Necurs are used to launch malware attacks in IoT communication environment. At the time of stealing mails from the user's mailbox, Emotet is applied. Emotet is capable to abduct the credentials of SMTP protocol, which is used for mail transfer. Gamut is good at for making spam e-mails. At preliminary stage, Gamut try to establish a communication with the target device. In order to launch new type of ransomware attack and different digital extortions, Necurs are used.

## 6. PERFORMANCE OF DIFFERENT MACHINE LEARNING CLASSIFIERS

Hackers are becoming very complicated and dangerous with the evolving technology and various new types of malware, making traditional methods of attack-prevention cumbersome. Therefore, protecting an IoT system or Cloud/Fog integrated IoT system becomes more challenging with the limited resources. To help detect these attacks, one of the widely used tools is machine learning (ML) algorithms. Several ML algorithms have proven extremely helpful in mitigating security as well as privacy attacks. The performance of various ML algorithms are summarized in Table 1. Some popular ML algorithms are random forest (RF), decision tree (DT), Naïve Bayes (NB), logistic regression (LR), K nearest neighbor classifier (kNN), support vector machine (SVM), linear discriminant analysis (LDA), etc. In Table 1, TPR and FPR means true positive rate and false positive rate, respectively.

The kNN machine learning classifier achieves better performance and accuracy in the detection of the malware where static features are considered.

**Table 1: Performance comparison of ML algorithms for various malware, intrusion and other types of attack detection**

| References | Purpose | Algorithm | Dataset | Results |
|---|---|---|---|---|
| [30] | Intrusion detection | KNN | Custom | Accuracy: 99.5% |
| | | LSVM | | Accuracy: 92.1% |
| | | DT | | Accuracy: 99.5% |
| | | RF | | Accuracy: 99.8% |
| | | NN | | Accuracy: 98.9% |
| [31] | Intrusion detection | RF | NSL-KDD | Accuracy for DoS attack: 99.67% |
| | | J48 | | Accuracy for DoS attack: 99.25% |
| [32] | Intrusion detection | DT | Custom | Accuracy for new attack profile: 64.66% |
| | | NB | | Accuracy for new attack profile: 57.38% |
| | | LDA | | Accuracy for new attack profile: 56.00% |
| [33] | Malware detection of Android system | DT | - | AUC: 96.4%, Accuracy: 95.4% |
| | | NB | - | AUC: 91.5%, Accuracy: 86.7% |
| | | LR | - | AUC: 97.7%, Accuracy: 93.2% |
| | | PART | - | AUC: 97.0%, Accuracy: 96.3% |
| [34] | Adeversial attack detection | RF | - | Accuracy: 92.79% (Features: Permission) |
| [35] | Adeversial attack detection | RF | - | Accuracy: 94.90% (Features: Permission) |
| [36] | Adeversial attack detection | RF | - | Accuracy: 92.36% (Features: Permission, API calls) |
| [37] | Adeversial attack detection | KNN | - | Accuracy: 97.87% (Features: Permission, API calls) |
| [38] | Impersonate attack detection | KNN | 3987 malware apps collected from different sources (McAfee and Android Malware Genome Project) | Accuracy: 99%, FPR: 2.2% (Features: API calls) |
| [39] | Android Malware detection | Improved Naïve Bayes | Collected from the Google Play Store and Chinese App store (6192 benign, 5560 malware apps) | TPR: 98.2%, FPR: 98.2%, Accuracy: 98% |
| | | NB | | TPR: 80.5%, FPR: 80.7%, Accuracy: 90.5% |
| | | SVM | | TPR: 95.2%, FPR: 95.2%, Accuracy: 95% |
| | | KNN | | TPR: 75.8%, FPR: 87.5%, Accuracy: 92% |
| [40] | Event-aware Android malware detection | Neural network | 10 956 benign samples in 2014 from PlayDrone [53], 4000 new apps in 2018 from Play Store [54], and 28 848 malicious samples from VirusShare [55]. | For 2014 & benign dataset: F1 score- 99.8 %, Precision-99.1%, Recall- 99. 2% <br> For 2018 & benign dataset: F1 score- 93.4 %, Precision-92.2%, Recall- 94.7% |
| [56] | Detection of new and unseen malicious applications | RIPPER | - | Accuracy: 89.36%, FPR: 7.77% |
| | | NB | - | Accuracy: 97.11%, FPR: 3.80% |
| | | Multi-NB | - | Accuracy: 96.88%, FPR: 6.01% |

## 7. RANSOMWARE IN IoT

Ransomware is a dangerous malware. It hijacks a user's system and steals all of his sensitive data. There are two types of ransomware: (a) Crypto-ransomware and, (b) Locker-ransomware. The Crypto-ransomware encrypts user's files and makes them inaccessible to the users [42]. The Locker-ransomware locks the user's device interface and demands for ransom to unlock the device. Recent ransomware attacks such as WannaCry and NotPetya have crushed the misconception that a back-up file can protect the digital data of an organization from being hacked [41]. If big institutions and firms were forced to pay money for unlocking file from ransom attack, one can only imagine the situation when a single individual is involved. In other words, when IoT, IoE and ransomware collide and cybercriminals begin to load IoT, IoE devices with the dangerous malware, a perfect storm of cyber security arms race will be created. When the malicious code is injected and circulated inside the system, the life cycle of ransomware starts at that time and lasts until the financial claim is shown to the victim. During this lifecycle, a number of activities are conducted in order to hijack the precious files and resources of the IoT user successfully. First, code dropper, mail attachment, or drive by download is utilized for smoothing the path of entering the ransomware's way into the victim's machine [51, 58]. When appeared, the malicious program in the host machine begins a flow of actions. These actions contain deleting shadow copies, creating a single computer ID, repairing the external IP address, etc., [57]. Ransomware interactions its command and control servers to achieve the encryption key. Then, in the next step, the malicious progress explores for user-related documents or files with particular extensions, i.e., .docx, .pptx, xlsx, .jpg, and .pdf. The encryption step holds place in the next step through passing the position of destined files or documents into a different position and then performing encryption on them. In this step, the encrypted documents or files are renamed. File extension of original files are changed and ransomware extensions are appended to files (e.g. locked, .crypto, _crypt, .crinf, .RDM, .RRK etc.) and the original files are deleted [57]. In the final stage, the financial claim is displayed by the malicious process in which ransom demands to the victim is included in the format of a text file or message of the desktop screen.

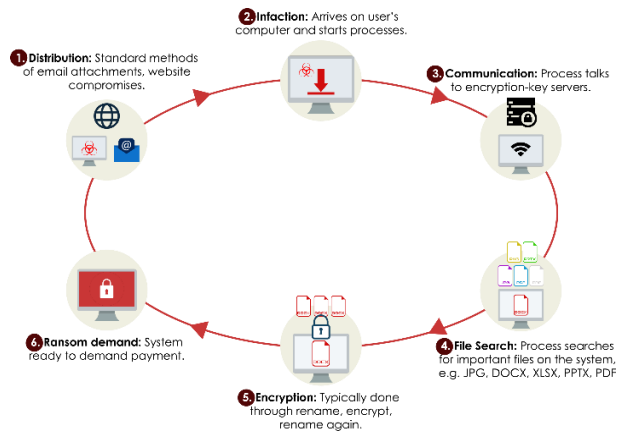

**Fig 9: Ransomware follows a number of typical steps to success [57]**

Al-rimy conducted a comprehensive survey and assessment of current ransomware related studies in his paper [43]. Even though these researches offered various solutions for ransomware recognition and prevention, there remains various open issues that require advance investigation and research. Here, this paper discusses these research directions and issues which can assist to develop the efficiency and effectiveness of ransomware recognition and prevention solutions [44]. Some existing software tools for detecting, analyzing and predicting ransomware are briefly illustrated in Table 2.

**Table 2: Tools for ransomware detection, classification and analysis**

| Operation Type | Methods | Reference paper |
|---|---|---|
| Static analysis | ApkTool. | [45], [46] |
| Dynamic analysis | Cuckoo Sandbox. | [47-50] |
| | Microsoft Filesystem Minifilter Driver. | [47] |
| Features extraction and factorization | Term Frequency-Inverse Document Frequency (TF-IDF), Term Frequency (TF), N-gram, Frequency-Centric Model (FCM), Natural Language Processing (NLP) | [45], [46], [50], [51] |
| Classification | SVM, LR, RF, Baysian Belief Network, NB. | [52], [49], [46], [50] |
| Similarity measurement | Structural similarity (SSIM), Cosine similarity. | [45], [47] |

## 8. FUTURE SCOPE AND RESEARCH DIRECTION

In the presence of the recent coronavirus disease 2019 (COVID-19) [62-63], [65-66], the importance of IoMT has greatly increased. Different machine learning [68-69] and deep learning techniques along with sensors [64], image processing [67] and wireless communication techniques [70-71] can be used to develop IoMT suitable for detecting and monitoring COVID-19 patients. In this section, several research challenges in hereafter, directions and scopes of malware detection in fog/cloud integrated or IoT related environment.

a) Robust security: Various recent malware detection and prevention methods do not provide full proof security against the new type of malware attacks. Moreover, some of these are attack specific and do not work for other types of attacks at the same time. Therefore, malware detection methods should be robust against multiple malware attacks at the same time.

b) Less computation time and less cost: IoMT/IoT communication environment consists of resource-constrained devices. This means that smart IoT devices have less storage capacity, short battery life, low computation power. Hence, we cannot use weighty deep learning systems for the malware detection for IoMT/IoT devices. As a result, malware detection methods and prevention mechanisms need to be designed in such a way that the proposed mechanisms must exhibit less computation time and less communication cost without negotiating the security needs.

c) Scalability: IoT is a type of heterogeneous network of different communication standards and applications. These applications have their own requirements and capabilities. In such an environment, we can have the "Electronic Health Records (EHRs)" of certain users that are stored in an IoT-enabled cloud server for further processing. The different devices inside the "Body Area Network (BANs)" produce data and send that to the cloud server. Therefore, it constructs a heterogeneous network of different communicating devices. We need a specific type of malware detection mechanism which can protect all types of devices of such kind of communication environment. Hence, more deep research study is needed in this direction.

d) Use of blockchain: The operations of blockchain can be conducted to secure various communication environments. It is because the blockchain operations are decentralized, efficient and transparent. Blockchain operations can also be utilized in efficient detection of the malware in IoT/IoMT environment. In such kind of detection method, we can create a block containing the information about the malicious programs (i.e., malware) to add in the blockchain. Since the blockchain is available to all authorized parties, these parties can have get into the information of the existing malware attacks.

## 9. CONCLUSION

With the advent of modern technologies, it is assumed that the number of smart devices will increase to a great extent. This will lead to the extension of IoT networks. IoT is being used in health care, industry as well as in day to day life. Particularly, IoT can be useful in managing current and future pandemics including COVID-19. With the increase in the number of interconnected devices and the growth of IoT, the number of security vulnerabilities are increasing. Security aspects are in the connected devices, in the data communication process, and in the data storage techniques. Security threats in the form of malware, ransomware, etc., may hinder the progress of IoT. Several security measures are being developed to ensure reliable IoT. This paper advances the current state of the art of security aspects of IoT. This work presents a number of potential security threats in the context of IoT, and discusses about the machine learning algorithms that can be useful in combating these threats. The findings of this paper will help develop more secure IoT networks and provide users secure user experience.